\documentclass[
    ,final            
  ]
  {aipproc}

\layoutstyle{6x9}

\usepackage{epsfig}
\usepackage{wrapfig,rotating}
\usepackage{amssymb}
\usepackage{amsmath}
\usepackage{hyperref}
\newcommand{\be}{\begin{equation}}
\newcommand{\ee}{\end{equation}}
\newcommand{\beq}{\begin{equation}}
\newcommand{\eeq}{\end{equation}}
\newcommand{\bea}{\begin{eqnarray}}
\newcommand{\eea}{\end{eqnarray}}

\newcommand{\half}{{\frac{1}{2}}}
\newcommand{\ket}[1]{\,\left|\,{#1}\right\rangle}
\newcommand{\mbf}[1]{\mathbf{#1}}

\begin{document}

\title{Light-Front Holography \\ and Hadronization at the Amplitude Level}


\author{Stanley J. Brodsky}{
  address={Stanford Linear Accelerator Center, 
Stanford University, Stanford, CA 94309, USA 
and the Institute for Particle Physics and Phenomenology, Durham, UK}
  }
  \author{Guy F. de T\'eramond}{
  address={Universidad de Costa Rica, San Jos\'e, Costa Rica}
  }
  \author{Robert Shrock}{
  address={C.N. Yang Institute for Theoretical Physics, Stony Brook University, 
Stony Brook, NY 11794, USA}
  }

\begin{abstract}

The correspondence between theories in anti--de Sitter space  and conformal field theories in physical space-time leads to an analytic, semiclassical model for strongly-coupled QCD which has scale invariance at short distances and color confinement at large distances.  Light-front holography is a remarkable feature of AdS/CFT:  it allows hadronic amplitudes in the AdS fifth dimension to be mapped to frame-independent light-front wavefunctions of hadrons in physical space-time, thus providing a  relativistic description of hadrons at the amplitude level.   Some novel features  of QCD are discussed, including the consequences of confinement for quark and gluon condensates and the behavior of the QCD coupling in the infrared.   We suggest that the spatial support of QCD condensates
is restricted to the interior of hadrons, since they arise due to the
interactions of confined quarks and gluons.  Chiral symmetry is thus broken in a limited domain of size $1/ m_\pi$,  in analogy to the limited physical extent of superconductor phases.
A new method for computing the hadronization of quark and gluon jets at the amplitude level, an event amplitude generator, is outlined. 

\end{abstract}

\maketitle
\section{Introduction}

One of the most challenging problems in strong interaction dynamics is to understand the interactions and composition of hadrons in terms of the fundamental quark and gluon degrees of freedom of the QCD Lagrangian. Because of the strong-coupling of QCD in the infrared domain, it has been difficult to find analytic solutions for the wavefunctions of hadrons or to make precise predictions for hadronic properties outside of the perturbative regime.  Thus an important theoretical goal is to find an initial approximation
to bound-state problems in QCD which is analytically tractable and which can be systematically improved.
Recently 
the AdS/CFT correspondence~\cite{Maldacena:1997re} between string
states in anti--de Sitter (AdS) space and conformal field theories in physical space-time, modified for color confinement,
has led to a semiclassical model for strongly-coupled QCD which provides analytical insights
into its inherently non-perturbative nature, including hadronic spectra, decay constants, and wavefunctions.

Conformal symmetry is broken in physical QCD by quantum effects and quark masses.
There are indications however,  both from
theory and phenomenology, that the QCD coupling is slowly varying at small momentum 
transfer~\cite{Brodsky:2008pg}. In particular, a new extraction of the effective strong coupling constant
$\alpha_s^{g_1}(Q^2)$
from CLAS spin structure function data in an extended $Q^2$ region
using the Bjorken sum rule $\Gamma_1^{p-n}(Q^2)$~\cite{Deur:2008rf},
indicates the lack of $Q^2$ dependence of $\alpha_s$ in the low $Q^2$ limit. One can understand this
physically: in a confining theory where gluons have an effective 
mass~\cite{Cornwall:1981zr} or the  quarks and gluons have 
maximal wavelength~\cite{Brodsky:2008be}, all vacuum polarization
corrections to the gluon self-energy decouple at long wavelength.  Thus an infrared fixed point appears to be a natural consequence of confinement.  Furthermore, if one considers a
semiclassical approximation to QCD with massless quarks and without
particle creation or absorption, then the resulting $\beta$ function
is zero, the coupling is constant, and the approximate theory is
scale and conformal invariant~\cite{Parisi:1972zy}. One can then use conformal symmetry as 
a {\it template}, systematically correcting for its nonzero $\beta$ function as well
as higher-twist effects~\cite{Brodsky:1985ve}.  In particular, we can use the mapping of the group of Poincare and conformal generators $SO(4,2)$ to the isometries of $AdS_5$ space.  The fact that gluons have maximum wavelengths
and hence minimum momenta provides an explanation for why multiple emission of
soft gluons with large couplings does not spoil the Appelquist-Politzer
argument for the narrowness of $J/\psi$ and $\Upsilon$; the explanation is that
such emission is kinematically forbidden~\cite{Brodsky:2008be}.

In AdS/CFT different values of the fifth dimension variable $z$ determine the scale of the invariant
separation between the partonic constituents. 
Hard scattering processes occur in the small-$z$ ultraviolet (UV)
region of AdS space. In particular,
the $Q \to \infty$ zero separation limit corresponds to the $z \to 0$ asymptotic boundary, where the QCD
Lagrangian is defined. 
In the large-$z$ infrared (IR) region  a cut-off is introduced to truncate the regime where the AdS modes can propagate. The infrared cut-off breaks conformal invariance, allows the introduction of
a scale and a spectrum of particle states. In the hard wall model~\cite{Polchinski:2001tt}
 a cut-off is placed at a
finite value $z_0 = 1/\Lambda_{\rm QCD}$ and the spectrum of states is linear in the radial and angular momentum quantum numbers:
$\mathcal{M} \sim 2 n \! + \! L$. In the soft wall model a smooth infrared cutoff is chosen to model
confinement and reproduce the usual Regge behavior 
$\mathcal{M}^2 \sim n \! + \!  L$~\cite{Karch:2006pv}.
The resulting models, although {\it ad hoc}, provide a
simple semiclassical approximation to QCD which has both constituent counting
rule behavior at short distances and confinement at large 
distances~\cite{Brodsky:2008pg,Polchinski:2001tt}.
It is thus natural, as a useful first approximation, to use the isometries of AdS to map the local interpolating operators at the UV boundary of AdS space to the modes propagating inside AdS.
The short-distance behavior of a hadronic state is
characterized by its twist  (dimension minus spin) 
$\tau = \Delta - \sigma$, where $\sigma$ is the sum over the constituent's spin
$\sigma = \sum_{i = 1}^n \sigma_i$. Twist is also equal to the number of partons $\tau = n$.
Under conformal transformations the interpolating operators transform according to their twist, and consequently the AdS isometries map the twist scaling dimensions into the AdS 
modes~\cite{Brodsky:2003px}.

As we have recently shown~\cite{Brodsky:2006uqa,Brodsky:2007hb}, there is a remarkable mapping between the AdS description of hadrons and the Hamiltonian formulation of QCD in physical space-time quantized on the light front.   The mathematical consistency of light-front holography for both the electromagnetic and gravitational~\cite{Brodsky:2008pf} hadronic transition matrix elements demonstrates that the mapping between the AdS holographic  variable $z$ and the transverse light-front variable $\zeta,$ which is a function of the multi-dimensional coordinates of the partons in a given light-front Fock state $x_i, \mbf{b}_{\perp i}$ at fixed light-front time $\tau,$ is a general principle.

When a flash from a camera illuminates a scene, each object is illuminated along the light-front of the flash; i.e., at a given $\tau$.  Similarly, when a sample is illuminated by an x-ray source such as the Linac Coherent Light  Source, each element of the target is struck at a given $\tau.$  In contrast, setting the initial condition using conventional instant time $t$ requires simultaneous scattering of photons on each constituent. 
Thus it is natural to set boundary conditions at fixed $\tau$ and then evolve the system using the light-front Hamiltonian $P^- = P^0-P^z = i {d/d \tau}.$  The invariant Hamiltonian $H_{LF} = P^+ P^- - P^2_\perp$ then has eigenvalues $\mathcal{M}^2$ where $\mathcal{M}$ is the physical mass.   Its eigenfunctions are the light-front eigenstates whose Fock state projections define the light-front wavefunctions.

The natural concept of a wavefunction for relativistic quantum field theories such as QCD is the light-front wavefunction $\psi_n(x_i, \mbf{k}_{\perp i}, \lambda_i)$ which specifies the  $n$ quark and gluon constituents of a hadron's Fock state as a function of the light-cone fractions $x_i = k^+/P^+ = (k^0+k^z)/(P^0+P^z)$ transverse momenta $\mbf{k}_{\perp i}$ and spin projections $\lambda_i$. The light-front wavefunctions of bound states in QCD are the relativistic
generalizations of the familiar Schr\"odinger wavefunctions of
atomic physics, but they are determined at fixed light-cone time
$\tau  = t +z/c$---the ``front form" advocated by Dirac~\cite{Dirac:1949cp}---rather than
at fixed ordinary time $t$.  Light-front wavefunctions are the fundamental process-independent amplitudes which encode hadron properties, predicting dynamical quantities such as spin correlations, form factors, structure functions, generalized parton distributions, and exclusive scattering amplitudes. 
The light-front wavefunctions  of a
hadron are independent of the momentum of the hadron, and they are
thus boost invariant;  Wigner transformations and Melosh rotations
are not required. The light-front formalism for gauge theories in
light-cone gauge is particularly useful in that there are no ghosts,
and one has a direct physical interpretation of  orbital angular
momentum.

{\it Light-Front Holography} is an important feature of AdS/CFT; it allows string modes $\Phi(z)$ in the AdS fifth dimension
to be precisely mapped to the light-front wavefunctions of hadrons in physical space-time in
terms of a specific light-front impact variable $\zeta$ which measures the separation of the
quark and gluonic  constituents  within the hadron.
The AdS/CFT correspondence implies
that a strongly coupled gauge theory is equivalent to the propagation
of weakly coupled strings in a higher dimensional space,
where physical quantities are computed in terms of an effective
gravitational theory. Thus, the AdS/CFT duality provides a gravity
description in a ($d+1$)-dimensional AdS
space-time in terms of a
$d$-dimensional conformally-invariant quantum field theory at the AdS 
asymptotic boundary~\cite{Gubser:1998bc}.

An important feature of light-front
quantization is the fact that it provides exact formulas for
current matrix elements as a sum of bilinear forms which can be mapped
into their AdS/CFT counterparts in the semiclassical approximation.
The AdS metric written in terms of light front coordinates $x^\pm =
x^0 \pm x^3$ is
\begin{equation} \label{eq:AdSzLF}
ds^2 = \frac{R^2}{z^2} \left( dx^+ dx^- - d \mbf{x}_\perp^2 - dz^2
\right).
\end{equation}
At fixed light-front time $x^+=0$, the metric depends only on the transverse
$ \mbf{x}_\perp$ and the holographic variable $z$.
Thus, as we show below, we can find an exact correspondence between the
fifth-dimensional coordinate of anti-de Sitter space $z$ and a
specific impact variable $\zeta$ in the light-front formalism which
measures the separation of the constituents within the hadron in
ordinary space-time.  The amplitude $\Phi(z)$ describing  the
hadronic state in $\rm{AdS}_5$ can then be precisely mapped to the valence
light-front wavefunctions $\psi_{n/H}$ of hadrons in physical
space-time~\cite{Brodsky:2006uqa,Brodsky:2007hb}, thus providing a relativistic
description of hadrons in QCD at the amplitude level.


Light-Front Holography can be derived by observing the correspondence between matrix elements obtained in AdS/CFT with the corresponding formula using the LF 
representation~\cite{Brodsky:2006uqa} .  The light-front electromagnetic form factor in impact 
space~\cite{Brodsky:2006uqa,Brodsky:2007hb,Soper:1976jc} can be written as a sum of overlap of light-front wave functions of the $j = 1,2, \cdots, n-1$ spectator
constituents:
\begin{equation} \label{eq:FFb} 
F(q^2) =  \sum_n  \prod_{j=1}^{n-1}\int d x_j d^2 \mbf{b}_{\perp j}  \sum_q e_q
\exp \! {\Bigl(i \mbf{q}_\perp \! \cdot \sum_{j=1}^{n-1} x_j \mbf{b}_{\perp j}\Bigr)} 
\left\vert \tilde \psi_n(x_j, \mbf{b}_{\perp j})\right\vert^2.
\end{equation}
The formula is exact if the sum is over all Fock states $n$.
For definiteness we shall consider a two-quark $\pi^+$  valence Fock state 
$\vert u \bar d\rangle$ with charges $e_u = \frac{2}{3}$ and $e_{\bar d} = \frac{1}{3}$.
For $n=2$, there are two terms which contribute to the $q$-sum in (\ref{eq:FFb}). 
Exchanging $x \leftrightarrow 1-x$ in the second integral  we find ($e_u + e_{\bar d}$ = 1)
\begin{eqnarray} \nonumber
 F_{\pi^+}(q^2)  &=& \!  \int_0^1 \! d x \int \! d^2 \mbf{b}_{\perp}  
 e^{i \mbf{q}_\perp \cdot  \mbf{b}_{\perp} (1-x)} 
\left\vert \tilde \psi_{u \bar d/ \pi}\! \left(x,  \mbf{b}_{\perp }\right)\right\vert^2 \\
\label{eq:PiFFb}
&=&  2 \pi \int_0^1 \! \frac{dx}{x(1-x)}  \int \zeta d \zeta\, 
J_0 \! \left(\! \zeta q \sqrt{\frac{1-x}{x}}\right) 
\left\vert\tilde\psi_{u \bar d/ \pi}\!(x,\zeta)\right\vert^2,
\end{eqnarray}
where $\zeta^2 =  x(1-x) \mathbf{b}_\perp^2$ and $F_\pi^+(q\!=\!0)=1$. 
Notice that by performing an identical calculation for the
$\pi^0$ meson the result is $F_{\pi^0}(q^2) = 0$ for any value of $q$, as expected
from $C$-charge conjugation invariance.

We now compare this result with the electromagnetic form-factor in AdS space:
\begin{equation} 
F(Q^2) = R^3 \int \frac{dz}{z^3} \, J(Q^2, z) \vert \Phi(z) \vert^2,
\label{eq:FFAdS}
\end{equation}
where $J(Q^2, z) = z Q K_1(z Q)$.
Using the integral representation of $J(Q^2,z)$
\begin{equation} \label{eq:intJ}
J(Q^2, z) = \int_0^1 \! dx \, J_0\negthinspace \left(\negthinspace\zeta Q
\sqrt{\frac{1-x}{x}}\right) ,
\end{equation} we can write the AdS electromagnetic form-factor as
\begin{equation} 
F(Q^2)  =    R^3 \! \int_0^1 \! dx  \! \int \frac{dz}{z^3} \, 
J_0\!\left(\!z Q\sqrt{\frac{1-x}{x}}\right) \left \vert\Phi(z) \right\vert^2 .
\label{eq:AdSFx}
\end{equation}
Comparing with the light-front QCD  form factor (\ref{eq:PiFFb}) for arbitrary  values of $Q$
\begin{equation} \label{eq:Phipsi} 
\vert \tilde\psi(x,\zeta)\vert^2 = 
\frac{R^3}{2 \pi} \, x(1-x)
\frac{\vert \Phi(\zeta)\vert^2}{\zeta^4}, 
\end{equation}
where we identify the transverse light-front variable $\zeta$, $0 \leq \zeta \leq \Lambda_{\rm QCD}$,
with the holographic variable $z$.

Matrix elements of the energy-momentum tensor $\Theta^{\mu \nu} $ which define the gravitational form factors play an important role in hadron physics.  Since one can define $\Theta^{\mu \nu}$ for each parton, one can identify the momentum fraction and  contribution to the orbital angular momentum of each quark flavor and gluon of a hadron. For example, the spin-flip form factor $B(q^2)$, which is the analog of the Pauli form factor $F_2(Q^2)$ of a nucleon, provides a  measure of the orbital angular momentum carried by each quark and gluon constituent of a hadron at $q^2=0.$   Similarly,  the spin-conserving form factor $A(q^2)$, the analog of the Dirac form factor $F_1(q^2)$, allows one to measure the momentum  fractions carried by each constituent.
This is the underlying physics of Ji's sum rule~\cite{Ji:1996ek}:
$\langle J^z\rangle = \half [ A(0) + B(0)] $,  which has prompted much of the current interest in 
the generalized parton distributions (GPDs)  measured in deeply
virtual Compton scattering. Measurements of the GDP's are of particular relevance
for determining the distribution of partons in the transverse
impact plane, and thus could be confronted with AdS/QCD predictions which follow
from the mapping of AdS modes to the transverse impact representation~\cite{Brodsky:2006uqa}.
An important constraint is $B(0) = \sum_i B_i(0) = 0$;  i.e.  the anomalous gravitomagnetic moment of a hadron vanishes when summed over all the constituents $i$. This was originally derived from the equivalence principle of gravity~\cite{Teryaev:1999su}.  The explicit verification of these relations, Fock state by Fock state, can be obtained in the light-front quantization of QCD in  light-cone 
gauge~\cite{Brodsky:2000ii}.  Physically $B(0) =0$ corresponds to the fact that the sum of the $n$ orbital angular momenta $L$ in an $n$-parton Fock state must vanish since there are only $n-1$ independent orbital angular momenta.

The light-front expression for the helicity-conserving gravitational form factor in impact space
is~\cite{Brodsky:2008pf}
\begin{equation} \label{eq:Ab}
A(q^2) =  \sum_n  \prod_{j=1}^{n-1}\int d x_j d^2 \mbf{b}_{\perp j}  \sum_f x_f
\exp \! {\Bigl(i \mbf{q}_\perp \! \cdot \sum_{j=1}^{n-1} x_j \mbf{b}_{\perp j}\Bigr)} 
\left\vert \tilde \psi_n(x_j, \mbf{b}_{\perp j})\right\vert^2,
\end{equation}
which includes the contribution of each struck parton with longitudinal momentum $x_f$
and corresponds to a change of transverse momentum $x_j \mbf{q}$ for
each of the $j = 1, 2, \cdots, n-1$ spectators. 
For $n=2$, there are two terms which contribute to the $f$-sum in  (\ref{eq:Ab}). 
Exchanging $x \leftrightarrow 1-x$ in the second integral we find 
\begin{eqnarray} \label{eq:PiGFFb} \nonumber
A_{\pi}(q^2) &\! = \!&  2 \! \int_0^1 \! x \, d x \int \! d^2 \mbf{b}_{\perp}  
 e^{i \mbf{q}_\perp \cdot  \mbf{b}_{\perp} (1-x)} 
\left\vert \tilde \psi \left(x, \mbf{b}_{\perp }\right)\right\vert^2 \\
&\! = \!& 4 \pi \int_0^1 \frac{dx}{(1-x)}  \int \zeta d \zeta\,
J_0 \! \left(\! \zeta q \sqrt{\frac{1-x}{x}}\right)  \vert\tilde\psi(x,\zeta)\vert^2,
\end{eqnarray}
where $\zeta^2 =  x(1-x) \mathbf{b}_\perp^2$.  We now consider the expression for the hadronic gravitational form factor in AdS space
\begin{equation} 
A_\pi(Q^2)  =  R^3 \! \! \int \frac{dz}{z^3} \, H(Q^2, z) \left\vert\Phi_\pi(z) \right\vert^2,
\end{equation}
where~\cite{Abidin:2008ku} $H(Q^2, z) = \half  Q^2 z^2  K_2(z Q)$.
The hadronic form factor is normalized to one at $Q=0$, $A(0) = 1$.
Using the integral representation of $H(Q^2,z)$
\begin{equation} \label{eq:intHz}
H(Q^2, z) =  2  \int_0^1\!  x \, dx \, J_0\!\left(\!z Q\sqrt{\frac{1-x}{x}}\right) ,
\end{equation}
we can write the AdS gravitational form factor 
\begin{equation} 
A(Q^2)  =  2  R^3 \! \int_0^1 \! x \, dx  \! \int \frac{dz}{z^3} \, 
J_0\!\left(\!z Q\sqrt{\frac{1-x}{x}}\right) \left \vert\Phi(z) \right\vert^2 .
\label{eq:AdSAx}
\end{equation}
Comparing with the QCD  gravitational form factor (\ref{eq:PiGFFb}) we find an  identical  relation between the light-front wave function $\tilde\psi(x,\zeta)$ and the AdS wavefunction $\Phi(z)$
in Eq. (\ref{eq:Phipsi}) obtained from the mapping of the pion electromagnetic transition amplitude.

\section{Holographic Light-Front Hamiltonian and Schr\"odinger Equation}

The above analysis provides an exact correspondence between the holographic variable $z$ and the
impact variable $\zeta$ which measures the transverse separation of the constituents within
a hadron.  The mapping of $z$ from AdS space to $\zeta$ in light-front frame  allows the equations of motion in AdS space to be recast in the form of  a
light-front Hamiltonian equation~\cite{Brodsky:1997de}
$
H_{LF} \ket{\phi} = \mathcal{M}^2 \ket{\phi},
$
a remarkable result which allows the discussion of the AdS/CFT solutions in terms of light-front equations in physical 3+1 space time.
By substituting $\phi(\zeta) =
\zeta^{-3/2} \Phi(\zeta)$ in the AdS scalar wave
equation 
we find an effective Schr\"odinger equation as a function of the
weighted impact variable $\zeta$~\cite{Brodsky:2006uqa,Brodsky:2007hb}
\begin{equation} \label{eq:Scheq}
\left[-\frac{d^2}{d \zeta^2} + V(\zeta) \right] \phi(\zeta) =
\mathcal{M}^2 \phi(\zeta),
\end{equation}
with the conformal potential $V(\zeta) \to - (1-4 L^2)/4\zeta^2$,
an effective two-particle light-front radial equation for mesons.
Its eigenmodes determine the hadronic mass spectrum. 
We have written above $(\mu R)^2 = - 4 + L^2$, where $\mu$ is the
five-dimensional mass  in the AdS wave equation.
The holographic hadronic light-front
wave function $\phi(\zeta) = \langle \zeta \vert \phi \rangle$  
represent the probability amplitude to find $n$-partons at
transverse impact separation $\zeta = z$.   Its
eigenvalues are determined by the boundary conditions at 
$\phi(z =1/\Lambda_{\rm QCD}) = 0$ and are given in terms of the roots of
Bessel functions: $\mathcal{M}_{L,k} = \beta_{L,k} \Lambda_{\rm
QCD}$. The normalizable modes are
\begin{equation}
\phi_{L,k}( \zeta) =   \frac{ \sqrt{2} \Lambda_{\rm QCD}}{J_{1+L}(\beta_{L,k})}
 \sqrt{\zeta} J_L \! \left(\zeta \beta_{L,k} \Lambda_{\rm QCD}\right)
 \theta\big(\zeta \le
\Lambda^{-1}_{\rm QCD}\big).
\end{equation}
The lowest stable state $L = 0$ is determined by the
Breitenlohner-Freedman bound~\cite{Breitenlohner:1982jf}.
Higher excitations are matched to the small $z$ asymptotic behavior of each string mode
to the corresponding
conformal dimension of the boundary operators
of each hadronic state. The effective wave equation
(\ref{eq:Scheq}) is a relativistic light-front equation defined at
$x^+ = 0$. The AdS metric $ds^2$ (\ref{eq:AdSzLF})  is invariant if
$\mbf{x}_\perp^2 \to \lambda^2 \mbf{x}_\perp^2$ and $z \to \lambda
z$ at equal light-front time  $x^+ = 0$. The Casimir operator for the rotation
group $SO(2)$ in the transverse light-front plane is $L^2$. This
shows the natural holographic connection to the light front.

The pseudoscalar meson interpolating operator
$\mathcal{O}_{2+L}= \bar q \gamma_5 D_{\{\ell_1} \cdots D_{\ell_m\}} q$, 
written in terms of the symmetrized product of covariant
derivatives $D$ with total internal space-time orbital
momentum $L = \sum_{i=1}^m \ell_i$, is a twist-two, dimension $3 + L$ operator
with scaling behavior determined by its twist-dimension $ 2 + L$. Likewise
the vector-meson operator
$\mathcal{O}_{2+L}^\mu = \bar q \gamma^\mu D_{\{\ell_1} \cdots D_{\ell_m\}} q$
has scaling dimension $2 + L$.  The scaling behavior of the scalar and vector AdS modes is precisely the scaling required to match the scaling dimension of the local pseudoscalar and vector-meson interpolating operators.    The spectral predictions for the hard wall model for both light meson and baryon states is compared with experimental data in~\cite{Brodsky:2008pg}.

\begin{figure}[!]
    \includegraphics[width=10cm]{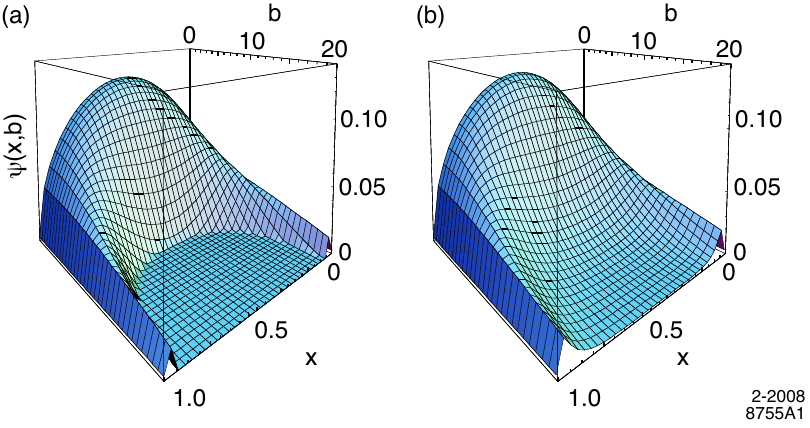}
  \caption{Pion light-front wavefunction $\psi_\pi(x, \mbf{b}_\perp$) for the  AdS/QCD (a) hard wall ($\Lambda_{QCD} = 0.32$ GeV) and (b) soft wall  ( $\kappa = 0.375$ GeV)  models.}
\label{fig2}  
\end{figure}

A closed form of the light-front wavefunctions $\tilde\psi(x, \mbf{b}_\perp)$ for the hard wall model follows from
(\ref{eq:Phipsi}) 
\begin{equation} 
\tilde \psi_{L,k}(x, \mbf{b}_\perp) 
=  \frac {\Lambda_{\rm QCD}}{\sqrt{ \pi} J_{1+L}(\beta_{L,k})} \sqrt{x(1-x)} 
J_L \! \! \left(\! \! \sqrt{x(1-x)} \, \vert\mbf{b}_\perp\vert \beta_{L,k} \Lambda_{\rm QCD}\!\right) 
\theta \! \Big(\mbf{b}_\perp^2 \le \frac{\Lambda^{-2}_{\rm QCD}}{x(1-x)}\Big).
\end{equation}
The resulting wavefunction (see fig. \ref{fig2})
displays confinement at large interquark
separation and conformal symmetry at short distances, reproducing dimensional counting rules for hard exclusive amplitudes.

\section{Spatially Restricted QCD Condensates}

Hadronic condensates play an important role in quantum chromodynamics (QCD).
Conventionally, these condensates are considered to be properties
of the QCD vacuum and hence to be constant throughout spacetime.
Recently  we have presented a new perspective on the nature of QCD
condensates $\langle \bar q q \rangle$ and $\langle
G_{\mu\nu}G^{\mu\nu}\rangle$, particularly where they have spatial and temporal
support~\cite{Brodsky:2008be,Brodsky:2008xm,Brodsky:2008xu}
We suggest that their spatial support is restricted to the interior
of hadrons, since these condensates arise due to the interactions of quarks and
gluons which are confined within hadrons.  Chiral symmetry is thus broken in a limited domain of size $1/ m_\pi$.  
For example, consider a meson consisting of a light quark $q$ bound to a heavy
antiquark, such as a $B$ meson.  One can analyze the propagation of the light
$q$ in the background field of the heavy $\bar b$ quark.  Solving the
Dyson-Schwinger equation for the light quark one obtains a nonzero dynamical
mass and, via the connection mentioned above, hence a nonzero value of the
condensate $\langle \bar q q \rangle$.  But this is not a true vacuum
expectation value; instead, it is the matrix element of the operator $\bar q q$
in the background field of the $\bar b$ quark.  The change in the (dynamical)
mass of the light quark in this bound state is somewhat reminiscent of the
energy shift of an electron in the Lamb shift, in that both are consequences of
the fermion being in a bound state rather than propagating freely.
Similarly, it is important to use the equations of motion for confined quarks
and gluon fields when analyzing current correlators in QCD, not free
propagators, as has often been done in traditional analyses of operator
products.  Since after a $q \bar q$ pair is created, the distance between the
quark and antiquark cannot get arbitrarily great, one cannot create a quark
condensate which has uniform extent throughout the universe.  The $55$ orders of magnitude conflict of QCD with the observed value of the cosmological condensate is thus removed~\cite{Brodsky:2008xu}.

This AdS/QCD model gives a good representation of the mass spectrum of
light-quark mesons and baryons as well as the hadronic wavefunctions.  One can also study the propagation of a scalar field $X(z)$
as a model for the dynamical running quark mass. The AdS
solution has the form~\cite{Babington:2003vm,Erlich:2005qh,DaRold:2005zs} $X(z) = a_1 z+ a_2 z^3$, where $a_1$ is
proportional to the current-quark mass. The coefficient $a_2$ scales as
$\Lambda^3_{QCD}$ and is the analog of $\langle \bar q q \rangle$; however,
since the quark is a color nonsinglet, the propagation of $X(z),$ and thus the
domain of the quark condensate, is limited to the region of color confinement.
The AdS/QCD picture of condensates with spatial support restricted to hadrons
is in general agreement with results from chiral bag models \cite{chibag},
which modify the original MIT bag by coupling a pion field to the surface of
the bag in a chirally invariant manner.  Since explicit breaking of ${\rm
SU}(2)_L \times {\rm SU}(2)_R$ chiral symmetry is small, and hence $m_\pi$ is
small relative to typical hadronic mass scales like $m_\rho$ or $m_N$, these
condensates can be treated as approximately constant throughout much of the
volume of a hadron. 

Our picture of confined condensates with spatial support restricted to the interior of
hadrons is consistent with the identification of pions as almost
Nambu-Goldstone bosons.  In our picture, the pions play a role analogous to the
Nambu-Goldstone modes, namely the quantized spin waves (magnons), that are
experimentally observed in a piece of a ferromagnetic substance below its Curie
temperature.  Again, strictly speaking, these spin waves result from the
spontaneous breaking of a continuous symmetry, which only occurs in an
idealized infinite-volume limit, but this limit provides a very good
approximation to a finite-volume sample.  The pions are the almost
Nambu-Goldstone bosons resulting from the spontaneous breaking of the global
${\rm SU}(2)_L \times {\rm SU}(2)_R$ chiral symmetry down to SU(2)$_{diag.}$,
and so, quite logically, the spatial support of their coordinate-space
wavefunctions is also a region where the chiral-symmetry breaking quark
condensate exists.  It is important to recall that the size of a hadron depends
not only on confinement but also on the virtual emission and reabsorption of
other hadrons, most importantly pions, since they are the lightest. Hence a
hadron can be regarded as being surrounded by a cloud of virtual pions.  By
general quantum mechanical arguments, this cloud is of size $\sim 1/m_\pi$. If
the two sources of explicit breaking of chiral ${\rm SU}(2)_L \times {\rm
SU}(2)_R$ symmetry were removed, i.e. the $m_u$ and $m_d$ current-quark masses
were taken to zero and the electroweak interactions were turned off, so that
$m_\pi=0$, then the size of a hadron, including its pion cloud, would increase
without bound until it impinged on neighboring hadrons.  In this case, the
quark and gluon condensates would also extend throughout all of spacetime.
Thus, our picture of condensates reduces to the conventional view in the chiral
limit.  

\section{Hadronization at the Amplitude Level}

\begin{figure}[!]
\includegraphics[width=10cm]{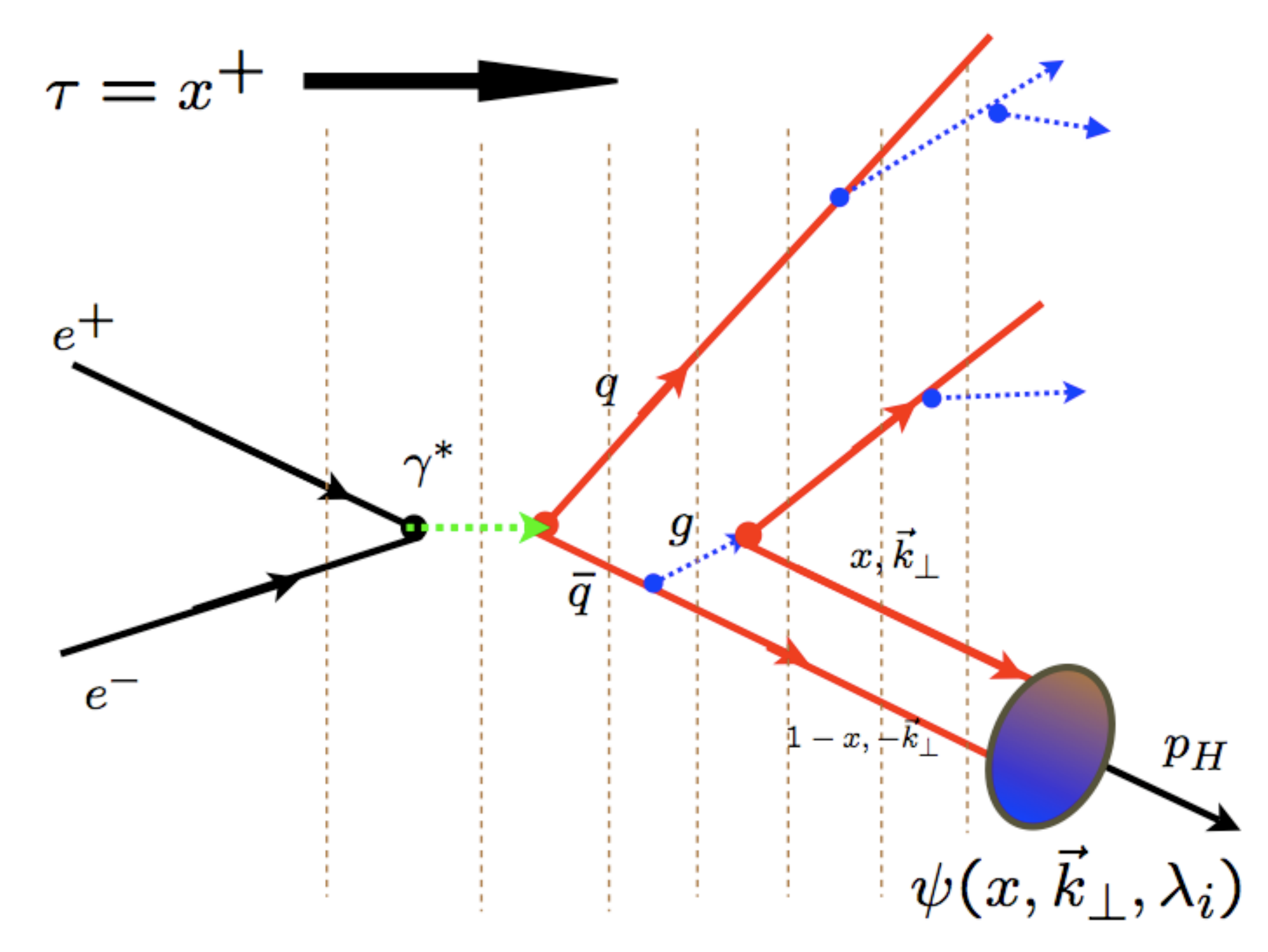}
  \caption{Illustration of an event amplitude generator for $e^+ e^- \to \gamma^* \to X$ for 
  hadronization processes at the amplitude level. Capture occurs if
  $\zeta^2 = x(1-x) \mbf{b}_\perp^2 > 1/ \Lambda_{\rm QCD}^2$
   in the AdS/QCD hard wall model of confinement;  i.e. if
  $\mathcal{M}^2 = \frac{\mbf{k}_\perp^2}{x(1-x)} < \Lambda_{\rm QCD}^2$.}
\label{fig1}  
\end{figure}

The conversion of quark and gluon partons is usually discussed in terms  of on-shell hard-scattering cross sections convoluted with {\it ad hoc} probability distributions. 
The LF Hamiltonian formulation of quantum field theory provides a natural formalism to compute 
hadronization at the amplitude level.  In this case one uses light-front time-ordered perturbation theory for the QCD light-front Hamiltonian to generate the off-shell  quark and gluon T-matrix helicity amplitude  using the LF generalization of the Lippmann-Schwinger formalism:
\begin{equation}
T ^{LF}= 
{H^{LF}_I } + 
{H^{LF}_I }{1 \over {\cal M}^2_{\rm Initial} - {\cal M}^2_{\rm intermediate} + i \epsilon} {H^{LF}_I }  
+ \cdots 
\end{equation}
Here   ${\cal M}^2_{\rm intermediate}  = \sum^N_{i=1} {(k^2_{\perp i} + m^2_i )/x_i}$ is the invariant mass squared of the intermediate state and ${H^{LF}_I }$ is the set of interactions of the QCD LF Hamiltonian in the ghost-free light-cone gauge~\cite{Brodsky:1997de}.
The $T^{LF}$-matrix element is
evaluated between the out and in eigenstates of $H^{QCD}_{LF}$.   The event amplitude generator is illustrated for $e^+ e^- \to \gamma^* \to X$ in fig. \ref{fig1}.

The LFWFS of AdS/QCD can be used as the interpolating amplitudes between the off-shell quark and gluons and the bound-state hadrons.
Specifically,
if at any stage a set of  color-singlet partons has  light-front kinetic energy 
$\sum_i {\mbf{k}^2_{\perp i}/ x_i} < \Lambda^2_{QCD}$, then one coalesces the virtual partons into a hadron state using the AdS/QCD LFWFs.   This provides a specific scheme for determining the factorization scale which  matches perturbative and nonperturbative physics.

This scheme has a number of  important computational advantages:

(a) Since propagation in LF Hamiltonian theory only proceeds as $\tau$ increases, all particles  propagate as forward-moving partons with $k^+_i \ge 0$.  There are thus relatively few contributing
 $\tau-$ordered diagrams.

(b) The computer implementation can be highly efficient: an amplitude of order $g^n$ for a given process only needs to be computed once.

(c) Each amplitude can be renormalized using the ``alternate denominator"  counterterm method~\cite{Brodsky:1973kb}, rendering all amplitudes UV finite.

(d) The renormalization scale in a given renormalization scheme  can be determined for each skeleton graph even if there are multiple physical scales.

(e) The $T^{LF}$-matrix computation allows for the effects of initial and final state interactions of the active and spectator partons. This allows for leading-twist phenomena such as diffractive DIS, the Sivers spin asymmetry and the breakdown of the PQCD Lam-Tung relation in Drell-Yan processes.

(f)  ERBL and DGLAP evolution are naturally incorporated, including the quenching of  DGLAP evolution  at large $x_i$ where the partons are far off-shell.

(g) Color confinement can be incorporated at every stage by limiting the maximum wavelength of the propagating quark and gluons.

A similar off-shell T-matrix approach was used to predict antihydrogen formation from virtual positron--antiproton states produced in $\bar p A$ 
collisions~\cite{Munger:1993kq}.

\section{Conclusions}

Light-Front Holography is one of the most remarkable features of AdS/CFT.  It  allows one to project the functional dependence of the wavefunction $\Phi(z)$ computed  in the  AdS fifth dimension to the  hadronic frame-independent light-front wavefunction $\psi(x_i, \mbf{b}_{\perp i})$ in $3+1$ physical space-time. The 
variable $z $ maps  to $ \zeta(x_i, \mbf{b}_{\perp i})$. To prove this, we have shown that there exists a correspondence between the matrix elements of the energy-momentum tensor of the fundamental hadronic constituents in QCD with the transition amplitudes describing the interaction of string modes in anti-de Sitter space with an external graviton field which propagates in the AdS interior. The agreement of the results for both electromagnetic and gravitational hadronic transition amplitudes provides an important consistency test and verification of holographic mapping from AdS to physical observables defined on the light-front. As we have discussed, this correspondence is a consequence of the fact that the metric $ds^2$ for AdS$_5$ at fixed light-front time $\tau$ is invariant under the simultaneous scale change  $\mbf{x}^2_\perp \to \lambda^2 \mbf{x}^2_\perp $ in transverse space and $z^2 \to \lambda^2 z^2$.  The transverse coordinate $\zeta$ is closely related to the invariant mass squared  of the constituents in the LFWF  and its off-shellness  in  the light-front kinetic energy,  and it is thus the natural variable to characterize the hadronic wavefunction.  In fact $\zeta$ is the only variable to appear in the light-front
Schr\"odinger equations predicted from AdS/QCD.  These equations for both meson and baryons give a good representation of the observed hadronic spectrum, especially in the case of the soft wall model. The resulting LFWFs also have excellent phenomenological features, including predictions for the  electromagnetic form factors and decay constants.  We have also shown that the LF Hamiltonian formulation of quantum field theory provides a natural formalism to compute 
hadronization at the amplitude level.

It is interesting to note that the form of the nonperturbative pion distribution amplitude $ \phi_\pi(x)$ obtained from integrating the $ q \bar q$ valence LFWF $\psi(x, \mbf{k}_\perp)$  over $\mbf{k}_\perp$,
has a quite different $x$-behavior than the
asymptotic distribution amplitude predicted from the PQCD
evolution~\cite{Lepage:1979zb} of the pion distribution amplitude.
The AdS prediction
$ \phi_\pi(x)  = \sqrt{3}  f_\pi \sqrt{x(1-x)}$ has a broader distribution than expected from solving the ERBL evolution equation in perturbative QCD.
This observation appears to be consistent with the results of the Fermilab diffractive dijet 
experiment~\cite{Aitala:2000hb}, the moments obtained from lattice QCD~\cite{Brodsky:2008pg} and pion form factor data~\cite{Choi:2006ha}.

Nonzero quark masses are naturally incorporated into the AdS predictions~\cite{Brodsky:2008pg} by including them explicitly in the LF kinetic energy  $\sum_i ( {\mbf{k}^2_{\perp i} + m_i^2})/{x_i}$. Given the nonpertubative LFWFs one can predict many interesting phenomenological quantities such as heavy quark decays, generalized parton distributions and parton structure functions.  
The AdS/QCD model is semiclassical and thus only predicts the lowest valence Fock state structure of the hadron LFWF.  In principle, the model can be systematically improved by diagonalizing the full QCD light-front Hamiltonian on the AdS/QCD basis.

The hard wall AdS/QCD model resembles  bag models where a boundary condition is introduced to implement confinement.  However, unlike traditional bag models, the AdS/QCD is frame-independent.  An important property of bag models is the  dominance of quark interchange as the underlying dynamics of large-angle elastic scattering,  This agrees with the survey of 
two-hadron exclusive reactions~\cite{White:1994tj}.  

Color confinement and its implementation in  AdS/QCD  implies a maximal
wavelength for confined quarks and gluons and thus a finite IR fixed point for
the QCD coupling. This strengthens our understanding of the narrow widths of
the $J/\psi$ and $\Upsilon$.   
We have also presented a new perspective on the nature of quark and gluon condensates in
quantum chromodynamics.  We suggest that the spatial support of QCD condensates
is restricted to the interior of hadrons, since they arise due to the
interactions of confined quarks and gluons~\cite{Brodsky:2008xm}.  Chiral symmetry is thus broken in a limited domain of size $1/ m_\pi$,  in analogy to the limited physical extent of superconductor phases.
Our picture explains the
results of recent studies which find no significant signal for the vacuum gluon
condensate.

\section*{Acknowledgments}
Presented by SJB at the 
Sixth International Conference on Perspectives in Hadronic Physics, 
12 - 16 May 2008, Miramare - Trieste, Italy.  He also thanks the Institute for Particle Physics Phenomenology, Durham, UK for its hospitality.
This research was supported by the Department
of Energy contract DE--AC02--76SF00515 and NSF-PHY-06-53342.
Pub numbers: SLAC-PUB-13306,  51DCPT/08/102,  YITP-SB-08-33.

\end{document}